\documentstyle[psfig,11pt,newpasp,twoside]{article}
\index{instructions}
\index{guidelines}

\def\edcomment#1{\iffalse\marginpar{\raggedright\sl#1\/}\else\relax\fi}
\reversemarginpar

\begin{document}
\title{Optical and Infrared Observations of Anomalous X-ray Pulsars}
\author{Martin Durant, Marten H van Kerkwijk}
\affil{University of Toronto, 60 St George St, Toronto, M5S 3H8, Canada}
\author{Ferdi Hulleman}
\affil{Universiteit Utrecht, Postbus 80000, 3508 TA Utrecht, Netherlands}

\begin{abstract}
The detection of optical/infrared counterparts to Anomalous X-ray
Pulsars (AXPs) has greatly increased our understanding of these
systems. Models for the AXP phenomenon were based upon their X-ray
emission, and all but the {\it magnetar} model made predictions for the
optical/infrared that have now been falsified.

With detections in hand, detailed studies of the optical/infrared to
X-ray flux ratios, variability, and the spectral energy distributions
have become possible. We present new data on two AXPs taken with
Keck and Magellan, and compare the results with predictions made
in the context of the magnetar model, in which the emission is due to
ion currents flowing in the $\ga10^{14}$G magnetosphere of young neutron stars.

\end{abstract}

\section{Introduction}
Anomalous X-ray Pulsars (AXPs) have been classified as a distinct category of 
neutron stars on the basis of observational properties which they
share in common. They were detected as X-ray sources with
long ($\sim$10s) pulse periods. The term {\it
anomalous} refers to the fact that the power associated with
rotational energy loss through spin-down is not sufficient to produce
the observed luminosities, nor were they obviously associated with
binary companions.

If they are short-lived, as supernova remnant associations suggest,
then AXPs along with the Soft Gamma-ray Repeaters (SGRs) and Dim
Isolated Neutron Stars (DINSs) hint at a large population of
neutron stars which differ considerably to the radio pulsars and X-ray
pulsars in binaries.

The earlies models were based purely on the X-ray properties of AXPs,
but with the identification and photometry of the first optical and
infrared counterparts, it became possible to test these models against
the measured flux ratio $F_X/F_{IR}$. 

Currently, the forerunner
to explain these objects is the {\it magnetar} model (Thompson \&
Duncan, 1996), in which AXPs are neutron stars whith 
dipole magnetic fields of order $B\ga 10^{14}$ Gauss, the decay of
which supplies the energy of the observed luminance. Thus they are
thought to be very similar objects to the SGRs when in quiescence,
although bursting activity has now been seen from AXPs also. This brings
into view the prospect of probing the physics of super-QED field
physics. One would expect to see variation in the optical or IR,
if the emission at those wavelengths is directly related to the processes causing the
bursting activity.

Since these first measurements, further optical and infrared
observations have been undertaken. It is now starting to be possible
to construct spectral energy distributions and to look for broad
features. In this way, we can gain more information and another handle
on these intriguing objects.

\section{AXP 4U 0142}
Optical and infrared images were taken of the field of AXP 4U 0142+61
using LRIS and NIRC at Keck respectively.

\begin{figure}[ht]
\begin{center}
\psfig{figure=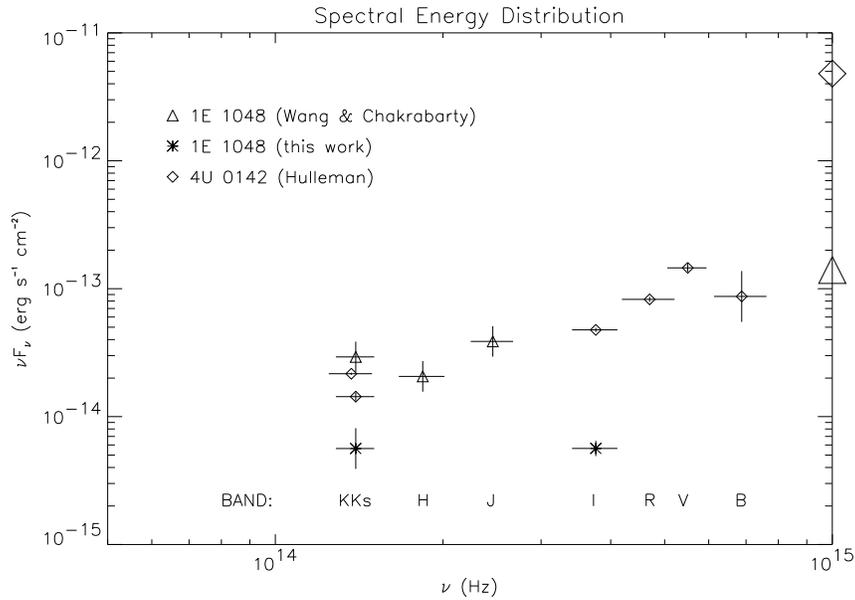,width=11.5cm}
\caption{The spectral energy distributions of the two AXPs in infrared
  and optical. The two larger symbols on the right-hand
  axis indicate the $\nu F_\nu / 10^2$ of the X-ray emission for each
  object. Not shown are Israel et al's limits for 1E 1048.}
\label{bigfig}
\end{center}
\end{figure}

The object was clearly seen in the V, R, I, K and Ks bands, but
after two hours of integration there
was only a marginal ($2\sigma$)  detection in the B band, which could be
seen either as a tentative detection or a hard upper limit in magnitude. On
the plot, Figure \ref{bigfig}, the spectral energy distribution (SED) of
4U0142 is shown from the infrared to the X-ray, de-reddened with an
$A_V=5.1$, the figure inferred from the X-ray column density of
hydrogen. 

The points to note are that the optical/IR spectral points do not
match   extrapolations of the blackbody + power law spectrum which was
fit to the X-ray (White et al, 1996); there appears to be a break in the continuum in B;
there is a marked difference between the K and Ks bands which, due to
their spectral proximity, is most easily attributed to variability
(since the respective measurements were taken over two and a half
years apart). Conversely, the optical flux appears to have been
stable. 

The
choice of reddening has a marked impact on the shape of the SED, but
the above three statements remain valid for the range of $A_V$ that
has been suggested, i.e. $A_V=2.6$ from a nearby OB association
(Hulleman et al, 2001) to $5.1$
from the Hydrogen column inferred from the X-ray spectrum (Patel et
al, 2002). Full details of these results are presented in Hulleman (2003).

\section{AXP 1E 1048}
\begin{figure}
\begin{center}
\psfig{figure=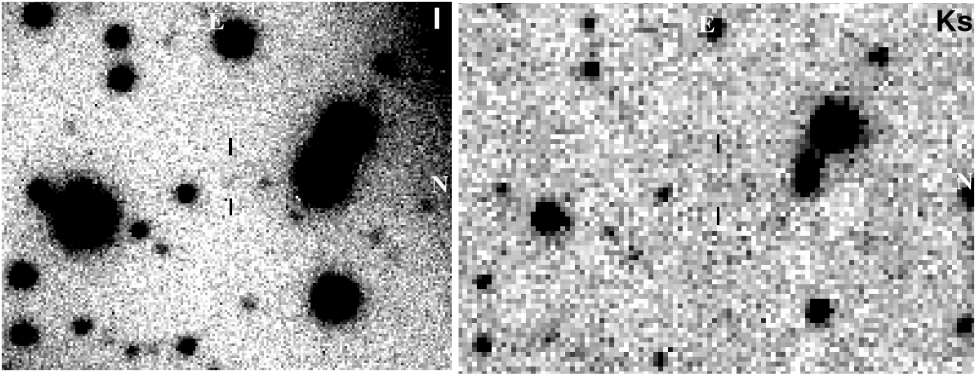,width=13cm}
\caption{I- and K$_{s}$-band image of 1E 1048, taken with MagIC and
  PANIC on Magellan II (Clay) at
  LCO. The AXP's position is indicated by white marks as found by
  e.g. Wang and Chakrabarty (2002). It is clearly visible, but very faint.}
\label{1048I}
\end{center}
\end{figure}

Optical and infrared images were taken of AXP 1E 1048.1-5937 using
MagIC and PANIC on the Magellan II (Clay) telescope, Las Campanas and
using ISAAC at VLT. See Figure \ref{1048I} for the I- and K$_{s}$-band images.
The analysis is still ongoing, and the results presented here are thus
preliminary.

The object was detected in the I band at an apparent  magnitude of
$I=26.6 \pm 0.15$ after over two hours' exposure time at superb seeing
conditions of $0\farcs3$. This may fairly be said to be at the
limits of what is possible at the current time. Also, it is worth
noting that our Ks band magnitude of $21.2 \pm 0.4$ is already fainter than
that which was measured by Wang \& Chakrabarty (2002), but consistent with the upper
limits measured by Israel et al (2002). Thus the hope is clearly that
here we are seeing the AXP in its quiescent state.

See again Figure \ref{bigfig} where the points for this AXP are also
plotted, de-reddened with $A_V=5.8$ (Wang \& Chakrabarty, 2002). We
see that the gradient of the spectrum from Ks to I is now different to
what was seen before, different to 4U 0142+61;  and that the infrared emission has varied
considerably.

\section{Implications}

The results above seem to suggest that the
optical emission seems stable whereas the
near infrared emission is highly variable. This in turn suggests 
that the source for the optical and the infrared emission is different,
and perhaps that the latter is related in some way to, or more
influenced by bursting activity.

The {\em Magnetar} model gives a qualitative explanation for the optical/IR
flux, in terms of ion currents flowing through the magnetosphere which
maintain the global twist in the magnetic field. If this is so, then
one would expect a relationship between the torque working on the star
and the optical flux. Thus the comparison of optical emission to the
spin down history of the AXPs will yield this relationship, and so it
is vital that both optical and X-ray monitoring are continued.

There is also the suggestion that the magnetar model may be able to
explain the spectral break (seen in the B band of 4U 0142) coming from
a particular radius in the magnetosphere, having a weak breaking
frequency dependence on the magnetic field ($\nu_{break} \propto B^{1/5}$). This
is can be verified with further photometry.

\end{document}